\documentclass[sigconf,table,xcdraw,dvipsnames]{acmart}

\usepackage{multirow}
\usepackage{xcolor}
\usepackage{graphicx}
\usepackage{dirtytalk}
\usepackage{hyperref}
\definecolor{Expert}{HTML}{FCE5CD}
\definecolor{Novice}{HTML}{C9DAF8}
\newcommand{\expert}[1]{\colorbox{Expert}{\hyperref[tab:participants]{E#1}}}
\newcommand{\novice}[1]{\colorbox{Novice}{\hyperref[tab:participants]{N#1}}}

\usepackage{colortbl}
\AtBeginDocument{%
  \providecommand\BibTeX{{%
    \normalfont B\kern-0.5em{\scshape i\kern-0.25em b}\kern-0.8em\TeX}}}

\setcopyright{none}
\copyrightyear{2024}
\acmYear{2024}
\acmDOI{10.1145/3613904.3642377}

\acmConference[CHI ’24]{Proceedings of the CHI Conference on Human Factors
in Computing Systems (CHI ’24)}{May 11–16, 2024, Honolulu, HI, USA}

\begin{document}

\title[Learning Programming of ABM with LLM Companions]{Learning Programming of Agent-based Modeling with LLM Companions: Experiences of Novices and Experts Using ChatGPT \& NetLogo Chat}

\author{John Chen}
\affiliation{%
  \institution{Northwestern University}
  \city{Evanston, IL}
  \country{United States of America}
}
\email{civitas@u.northwestern.edu}

\author{Xi Lu}
\affiliation{%
  \institution{University of California, Irvine}
  \city{Irvine, CA}
  \country{United States of America}
}
\email{xlu30@uci.edu}

\author{David Du}
\affiliation{%
  \institution{Northwestern University}
  \city{Evanston, IL}
  \country{United States of America}
}
\email{duyuzhou2013@gmail.com}

\author{Michael Rejtig}
\affiliation{%
  \institution{University of Massachusetts Boston}
  \city{Boston, MA}
  \country{United States of America}
}
\email{michael.rejtig001@umb.edu}

\author{Ruth Bagley}
\affiliation{%
  \institution{Northwestern University}
  \city{Evanston, IL}
  \country{United States of America}
}
\email{ruth.bagley@northwestern.edu}

\author{Michael S. Horn}
\affiliation{%
  \institution{Northwestern University}
  \city{Evanston, IL}
  \country{United States of America}
}
\email{michael-horn@northwestern.edu}

\author{Uri J. Wilensky}
\affiliation{%
  \institution{Northwestern University}
  \city{Evanston, IL}
  \country{United States of America}
}
\email{uri@northwestern.edu}

\renewcommand{\shortauthors}{Chen, et al.}

\begin{abstract}
Large Language Models (LLMs) have the potential to fundamentally change the way people engage in computer programming. Agent-based modeling (ABM) has become ubiquitous in natural and social sciences and education, yet no prior studies have explored the potential of LLMs to assist it. We designed NetLogo Chat to support the learning and practice of NetLogo, a programming language for ABM. To understand how users perceive, use, and need LLM-based interfaces, we interviewed 30 participants from global academia, industry, and graduate schools. Experts reported more perceived benefits than novices and were more inclined to adopt LLMs in their workflow. We found significant differences between experts and novices in their perceptions, behaviors, and needs for human-AI collaboration. We surfaced a knowledge gap between experts and novices as a possible reason for the benefit gap. We identified guidance, personalization, and integration as major needs for LLM-based interfaces to support the programming of ABM.
\end{abstract}



\received{14 September 2023}
\received[revised]{12 December 2023}
\received[accepted]{19 January 2024}

\maketitle

\section{Introduction}
The advent of coding-capable Large Language Models (LLMs) has the potential to fundamentally change the way people engage in computer programming\cite{eloundou_gpts_2023}. As LLM-based programming interfaces (e.g. GitHub Copilot; ChatGPT) become increasingly popular\cite{lau__2023}, some studies started to study their user perceptions\cite{vaithilingam_expectation_2022}. However, the research on their potential learning impacts is still limited. Many prior studies only focus on impressions of educators\cite{lau__2023} or students\cite{yilmaz_augmented_2023}, with little empirical data on the actual learning usage of these tools. On the other hand, a few studies started to explore how LLM-based interfaces can be designed to facilitate programming education, indicating potential advantages for learners. Notably, these studies suggest that learners with more prior programming experience tend to benefit more\cite{nam_-ide_2023, kazemitabaar_studying_2023}. While a recent study identifies some challenges for novice learners with LLM-based interfaces\cite{zamfirescu-pereira_why_2023}, there is a gap in understanding why experienced programmers seem to gain more learning benefits from these tools. 

In this paper, we present the design of a novel LLM-based interface, NetLogo Chat, for the learning and practice of NetLogo. NetLogo is a widely used programming language for agent-based modeling (ABM), which applies simple rules on multiple individual agents to simulate complex systems\cite{wilensky_netlogo_1997}. It is particularly powerful in capturing emergent phenomena, e.g., the spread of viruses or predator-prey systems\cite{wilensky_introduction_2015}. It is an important methodology in computational modeling across scientific disciplines and education from K-12 to postgraduate levels\cite{weintrop_defining_2016}, where scientists and educators are highly in need of LLM-based interfaces\cite{pal_domain-specific_2023, cooper_examining_2023}. As an important part of computational modeling, the priorities of ABM differ from general programming\cite{pylyshyn_computational_1978}. A modeler needs to verify that their conceptual design of individual rules matches the real-world patterns (e.g. a predator needs food to survive), the code matches the design (i.e. there are no unexpected or implicit assumptions), and the aggregated outcome matches real-world phenomena (e.g. if all prey die out, predators die too)\cite{fleischmann_ensuring_2009}. As most LLM-related studies on computer programming work on general-purpose languages that LLMs perform best (e.g. Python or Javascript), no LLM-related studies have explored ABM or other forms of computational modeling at this point.

NetLogo Chat was designed with constructionist learning principles and incorporated known best practices for ABM and computer programming. Constructionism advocates for the design of learning experiences where learners construct their understanding of the world (e.g. knowledge of ABM) through building personally meaningful artifacts (e.g. an agent-based model around learners' interests)\cite{papert_situating_1991}. Similar to GitHub Copilot Chat\cite{noauthor_using_nodate}, NetLogo Chat was integrated into an integrated development environment (IDE). Different from previous designs, it aims to give users more control over the human-AI collaboration processes, strives to incorporate authoritative sources, and tries to provide more support for troubleshooting. 

Using both ChatGPT and NetLogo Chat as a probe\cite{zamfirescu-pereira_why_2023}, we conducted a qualitative study to highlight the different perceptions, behaviors, and needs of experts and novices during open-ended modeling sessions. We interviewed 30 expert and novice participants from academia, industry, and graduate schools around the world. Participants proposed diverse NetLogo tasks from their disciplines and worked toward their modeling goals. We asked interview questions before, during, and after their interaction with each design. We answered the research questions:
\begin{enumerate}
  \item What perceptions - strengths, weaknesses, and adoption plans - do expert and novice users perceive LLM-driven interfaces to support their NetLogo learning and practice?
  \item How do expert and novice users use LLM-driven interfaces to support their NetLogo learning and practice? 
  \item What are expert and novice users’ needs for LLM-based interfaces to support their NetLogo learning and practice?
\end{enumerate}

Learners generally agreed with our design principles and suggested additional features for future designs. As in other studies, experts reported more perceived benefits than novices. Comparing the different interaction patterns between experts and novices, our study reveals a behavioral gap that might explain the gap in benefits. We found that experts collaborated with LLM-based interfaces with more human judgment in all activities than novices, helping them overcome AI hallucinations, while novices struggled with evaluating and debugging AI responses. From there, we identified components of a knowledge gap between novices and experts. We reported experts’ and novices’ needs in LLM-based interfaces in three key themes: guidance (from LLMs); personalization (of LLMs); and integration (into modeling environments), many of which confirm and develop the design decisions of NetLogo Chat. The contributions of this paper include: 

\begin{enumerate}
  \item The design and implementation of NetLogo Chat, an LLM-based system that supports learning and practice of NetLogo, a widely-used programming language for ABM;
  \item An empirical study that contributes to the understanding of how novices and experts perceive, use, and express needs for LLM-based programming interfaces in different ways;
  \item A theorization of the knowledge gap between experts and novices that might lead to the behavioral gap, and suggestions of potential design interventions;
  \item The design discussion and suggestions for building LLM-based programming interfaces that benefit both experts and novices in agent-based modeling more equitably.
\end{enumerate}

\section{Related Work}
\subsection{LLMs for Computational Programming and Modeling}
Researchers have been exploring natural-language-based interfaces for programming for decades, yet early attempts were mostly exploratory, being limited in capabilities. NaturalJava\cite{price_naturaljava_2000} required users to follow a strict pattern when prompting, while later systems (e.g. NaLIX\cite{li_nalix_2005} or Eviza\cite{setlur_eviza_2016}) asked for a specific set of English expressions. This created difficulties for users and system designers, as they felt “a main challenge of NLP interfaces is in communicating to the user what inputs are supported.”\cite{setlur_eviza_2016} Without the capability to generate natural languages, those interfaces were also constrained to one-off interactions.

Recently, a new generation of LLMs demonstrated the capability to understand and generate both natural languages and computer languages. GPT-3 was examined in writing code explanations\cite{macneil_generating_2022}, documentation\cite{khan_automatic_2022}, and providing feedback for assignments\cite{balse_investigating_2023}. Soon, educators started to believe that Codex could be used to solve simple programming problems\cite{finnie-ansley_robots_2022, wermelinger_using_2023}. Embedded in ChatGPT, GPT-3.5-turbo and GPT-4 demonstrated even stronger capabilities in programming. More and more LLMs have started to gain the capability of coding (e.g. PALM 2; Claude 2; CodeLLaMA 2), ushering in a new era of natural language interfaces for programming.

Even the most powerful LLMs suffer from hallucinations and may misunderstand human intentions. Early users of ChatGPT complained about incorrect responses and struggled to prompt ChatGPT for a desired output\cite{skjuve_user_2023}. While LLMs might outperform average humans in specific, structured tasks\cite{openai_gpt-4_2023}, the evaluation criteria might have been flawed\cite{liu_is_2023}, as LLMs struggled to combine existing solutions for a novel challenge\cite{dakhel_github_2023}. A study suggested that developers should not rely on ChatGPT when dealing with new problems \cite{tian_is_2023}.

LLMs are naturally less prepared in low-resource programming languages (LRPL). Here, our working definition for LRPL is similar to that of natural languages: with relatively scarce online resources and have been less studied by the AI field\cite{magueresse_low-resource_2020}. LRPLs are not less important: NetLogo, the most widely used programming language for agent-based modeling (ABM)\cite{thiele_agent-and_2011}, is used by hundreds of thousands of scientists, educators, and students for computational modeling. Using simple computational rules for individual agents, ABM could simulate complicated emergent phenomena. It has been frequently used in different scientific disciplines\cite{wilensky_introduction_2015} and science education\cite{hutchins_domain-specific_2020} for recent decades. With considerably fewer online resources to train on, LLMs are much more prone to errors and/or hallucinations with LRPLs\cite{tarassow_potential_2023}. 

A few studies attempted to improve LLMs' performance with LRPLs in two directions. First, some studies fine-tuned foundational LLMs with LRPL datasets\cite{chen_transferability_2022}. While this approach demands considerable datasets and computational power, it has not been applied to generative tasks yet\cite{gong_multicoder_2022}. Second, some studies used prompt engineering techniques. For example, aiming at simple tasks, a study creates sets of grammar rules for LLMs to fill in\cite{wang_grammar_2023}. Another study leveraged compiler outputs, allowing LLMs to iteratively improve their Rust code, but was only tested in a smaller number of fixed tasks\cite{wu_rustgen_2023}. The potential of LLMs in scientific disciplines, including in computational modeling, is rarely explored. At this point, the only study targeted at STEM helps with a very specific engineering task \cite{kumar_mycrunchgpt_2023}.

\subsection{User Perception and Behaviors with LLM-based Programming Interfaces}
Two strands of user perception and behaviors studies informed our design and study: studies of conversational agents (CAs); and of LLM-based programming interfaces. For education, CAs were used to develop learners’ writing\cite{wambsganss_arguetutor_2021}, self-talk\cite{fu_self-talk_2023}, and programming skills\cite{winkler_sara_2020}. Many of them are pedagogical conversational agents (PCA) with the aim to adaptively mimic the behaviors of human tutors\cite{winkler_unleashing_2018}. PCAs could serve in multiple roles, such as tutors\cite{wambsganss_arguetutor_2021}, motivators\cite{caballe_conversational_2019}, peer players\cite{gero_mental_2020}, or learning companions\cite{fu_self-talk_2023}.

Prior research of CAs underscored the importance of understanding user perception and behaviors\cite{gero_mental_2020}, yet the technical boundaries of the pre-LLM era limited the freedom of designers. Previous studies have explored aspects such as trust, mutual understanding, perceived roles\cite{clark_rethinking_2009}, privacy\cite{sannon_i_2020}, human-likeness\cite{jeong_exploring_2019}, utilitarian benefits, and user-related factors\cite{ling_factors_2021} to understand users’ acceptance and willingness to use CAs. However, many CAs before LLMs had to use pre-programmed responses\cite{wang_towards_2021}, and simply emulating functional rules from human speech failed to deliver people’s high expectations of CAs\cite{clark_what_2019}. Without the capability to read or write code, pre-LLM CAs for computing education were largely limited to providing relevant knowledge\cite{winkler_sara_2020} or supporting conceptual understanding of programming\cite{lin_zhorai_2020}.

Recent studies have started to understand user perception and behaviors with LLM-based programming interfaces. In education, early studies focused on instructors' and students’ perceptions of LLM-based interfaces for programming. Computer science students self-reported many potential benefits of using ChatGPT and were less inclined to report potential drawbacks\cite{yilmaz_augmented_2023}. On the other hand, computer science instructors were significantly concerned over students' widespread usage of ChatGPT\cite{lau__2023}. While some instructors went as far as banning ChatGPT altogether, others suggested exposing students to the capabilities and limitations of AI tools, leveraging mistakes in generated code for learning opportunities. Both instructors and students expressed the need to adapt to a new, LLM-era way of teaching and learning\cite{zastudil_generative_2023}. 

For professionals, challenges and opportunities co-exist with LLM-based programming interfaces. Recent studies found programmers preferred to use Copilot\cite{vaithilingam_expectation_2022} and finished tasks faster with Copilot\cite{peng_impact_2023}. Yet, Copilot struggled with more complicated problems, providing buggy or non-reproducible solutions\cite{dakhel_github_2023}. Professional programmers faced difficulties in understanding and debugging Copilot-generated code, which hinders their task-solving effectiveness\cite{vaithilingam_expectation_2022}. Programmers who trusted AI were prone to write insecure code with AI\cite{perry_users_2022}. For conversational interfaces, despite inputs being in natural languages, users felt that they needed to learn LLM’s “syntax”\cite{jiang_discovering_2022, fiannaca_programming_2023}. 

Our understanding of user perception and behaviors with LLM-based interfaces during (the learning of) computer programming is still very limited. As the field just started exploring this direction, previous studies mostly focused on general user impressions\cite{zastudil_generative_2023}, or conducted behavioral tasks on pre-scripted, close-ended tasks\cite{peng_impact_2023}. While close-ended settings made it easier to assess objective metrics\cite{blikstein_using_2011}, open-ended contexts open a wider window to understanding users’ learning patterns, behaviors, perceptions, and preferences\cite{blikstein_programming_2014}. For example, a recent study observed two modes that professional programmers interact in open-ended tasks with Copilot: acceleration, where the programmer already knows what they want to do next; and exploration, where the programmer uses AI to explore their options\cite{barke_grounded_2023}. Another study on professionals’ prompt engineering shed light on their struggles, challenges, and potential sources of behaviors\cite{zamfirescu-pereira_why_2023}. 

Still, we noticed two gaps in previous studies. First, a majority of studies chose professional programmers or computer science instructors/students as participants, while LLM-based interfaces are also used by millions of people without a CS background for programming tasks. Second, as HCI studies mostly focus on languages that LLMs are known to perform best, e.g. Python or HTML, little is known about user perceptions and behaviors when computational modeling or LRPLs are involved.

\subsection{LLM-based Interfaces for Learning Programming and Modeling}
While LLMs have shown promising potential in supporting human-AI collaboration in programming, most design studies were preliminary, and LLM-based interfaces for computational modeling remained understudied. For example, the Programmer’s Assistant integrated a chat window into an IDE\cite{ross_programmers_2023}. Going beyond simple integrations, GitHub Copilot Chat\cite{noauthor_using_nodate} provided in-context support within code editors, yet its user studies were still preliminary\cite{bull_generative_2023}. A similar design was done on XCode without a user study\cite{tan_copilot_2023}. Another study explored the integration between computational notebooks with LLMs and emphasized the role of the domain (in this case, data science) on LLM-based interface design\cite{mcnutt_design_2023}. 

LLMs have gained much attention among programming educators, but the design study is insufficient. Recent studies tested LLMs on introductory programming tasks and achieved unsurprisingly high scores\cite{savelka_thrilled_2023, chen_beyond_2023}. This prospect leads to great concerns among computer science instructors as they observed the widespread usage of ChatGPT among students\cite{lau__2023}. Yet, only a few LLM-based design studies targeted programming learning. Using a Wizard of Oz prototype, a study underscored the importance of supporting students' varied degrees of prior expertise\cite{robe_designing_2022}. A design study reported positive short-term performance gains when young, novice programming learners engaged with Codex\cite{kazemitabaar_studying_2023}. Another study also found LLMs’ benefits for novice programmers\cite{nam_-ide_2023}. Both studies found that more experienced programmers tended to benefit more, yet the reason was still unclear.

\label{Constructionism}
In this study, we invoke the learning theory of Constructionism\cite{papert_situating_1991} to inform our LLM-based system and empirical study design. While there is no rigid definition for Constructionism, it argues that learning happens most felicitously when learners "consciously engage in constructing a public entity"\cite{papert_situating_1991}. In the context of computer programming, it means learning happens naturally through programming computers, as it iteratively externalizes learners' internal understanding of the world in code, and then allows learners to improve their understanding through watching how the code runs\cite{papert_mindstorms_1980}. Moreover, it argues that computer programming is not as abstract or formal as it appears; individual programmers’ approaches are often concrete and personal, in pluralistic ways\cite{turkle_epistemological_1990}. However, the pluralism in thoughts is more difficult to capture by close-ended tasks (such as a problem set) and objective metrics (such as completion rate/time)\cite{blikstein_programming_2014}. As such, constructionist learning studies often prefer open-ended tasks (e.g. making games\cite{kafai_constructionist_2015}, designing instructional software\cite{harel_software_1990}, creating agent-based models in NetLogo\cite{blikstein_programming_2014}) and qualitative studies, as they open windows into the nuances of learners’ perceptions and behaviors in more natural and realistic settings.

The Logo programming language and its descendants (e.g. Scratch; Alice; NetLogo) succeeded in supporting multiple ways of knowing and thinking in computing education and in scientific research\cite{solomon_history_2020}, yet to our knowledge, no published studies have explored their synergy with LLMs. Many prominent constructionist design principles could be applied to AI-based interfaces\cite{kahn_constructionism_2021} and inspired the design of NetLogo Chat. For instance, “low floor, high ceiling, wide walls” asks learning environments to provide 1) an easy entrance for novices (low floor); 2) the possibility for experts to work on sophisticated projects (high ceiling); 3) the support of a wide range of different explorations (wide walls); 4) the support of many learning paths and styles\cite{resnick_reflections_2005}. We also learned from previous design studies that stress the importance of adaptive scaffolding\cite{chen_pocketworld_2023, sengupta_integrating_2013} and support debugging\cite{brady_debugging_2020} for novices to learn NetLogo. Hence, we contributed to the field one of the first design studies of LLM-based interfaces for learning programming that follow the constructionist tradition.

\section{NetLogo Chat System}
NetLogo Chat is an LLM-based system for learning and programming with NetLogo. It comprises two main parts: a web-based interface integrated with Turtle Universe (a version of NetLogo)\cite{chen_turtle_2021} (See \ref{Design Overview}); and an LLM-based workflow that improves the quality of AI responses and powers the interface (See \ref{Technical Implementation}). We iteratively designed the system by:

\begin{enumerate}
  \item Based on authors’ experiences in teaching NetLogo, we created a design prototype based on the constructionist learning theory (see \ref{Constructionism}), with a focus on supporting users iteratively build up their prompts and smaller code snippets before working on entire models. We developed a proof-of-concept system, using prompt engineering techniques to interact with GPT-3.5-turbo-0314. 
  \item We internally evaluated the proof-of-concept with a group of NetLogo experts. During this process, we encountered frequent hallucinations with NetLogo (grammatical or conceptual mistakes; inventing keywords that do not exist; etc). For the system to provide guidance, we realized that authoritative sources are necessary for LLMs’ performance;
  \item We incorporated the official NetLogo documentation and code examples into the system using prompt engineering techniques (see \ref{Technical Implementation}), evaluated other LLMs’ potential, and then conducted pilot interviews to evaluate the system with three external NetLogo experts invited from NetLogo's mailing lists. The interviews used a protocol similar to the one we formally used (see \ref{Interviews}), with more flexibility and open-endedness;
  \item Based on the external feedback, we identified the need for supporting troubleshooting, leading to the design decision \ref{Troubleshooting}. We upgraded the underlying LLM to GPT-3.5-turbo-0613, fixed many minor usability issues, and finalized the prototype that we used in the empirical study.
\end{enumerate}

\subsection{Design Overview}
\label{Design Overview}
\begin{figure}[h]
    \centering
    \includegraphics[width=240pt]{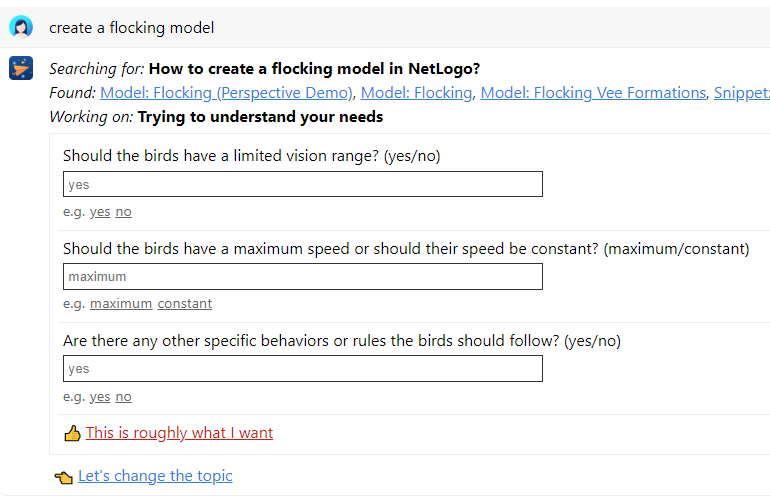}
    \caption{NetLogo Chat asking for details about human's needs.}
    \label{fig:NetLogo Chat}
    \Description{In this figure, the user asked NetLogo Chat "create a flocking model". NetLogo Chat started by searching for related documentation, provided several example models, then asked three follow-up questions to clarify the needs of the user.}
\end{figure}

\begin{figure}[h]
    \centering
    \includegraphics[width=240pt]{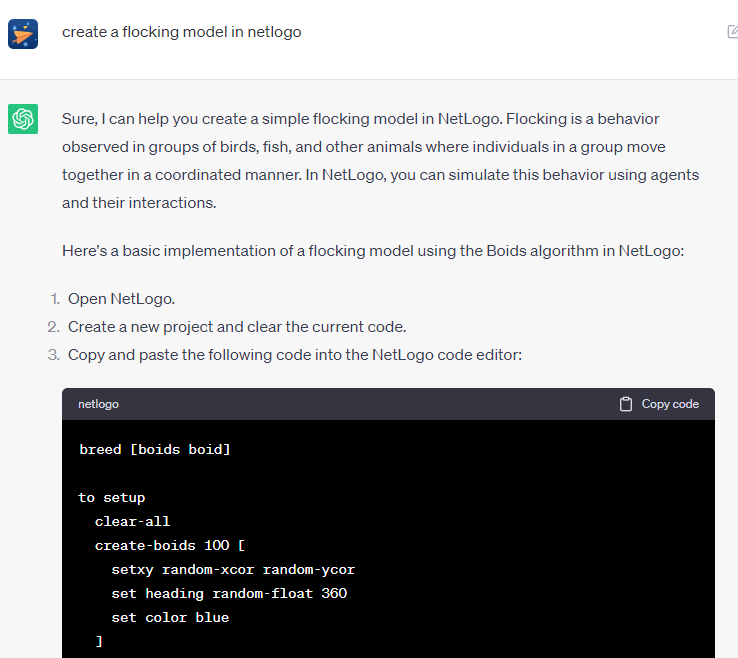}
    \caption{ChatGPT assuming details of human's needs.}
    \label{fig:chatgpt}
    \Description{In this figure, the user asked ChatGPT to "create a flocking model in netlogo". ChatGPT gave a direct response starting with "Sure, I can help you create a simple flocking model in NetLogo", then gave the user an instruction to open NetLogo, create a model, and copy a code snippet.}
\end{figure}

\subsubsection{Enable users to program the computer, rather than being programmed by the computer}
\label{Program}
Over-reliance on LLM-based interfaces has become a major concern among both educators and some learners, where students blindly follow the instructions given by LLMs without attempting to construct their representations of knowledge. Such a scenario is antithetical to the constructionist learning tradition, where Seymour Papert's fear of "computers program children" comes back to life again\cite{papert_mindstorms_1980}. 

Inspired by the Logo language, the design of NetLogo Chat aims to give control back to learners: to suppress LLMs' tendency to give a quick response that often assumes too much about the learner's inclination, we force it to ask clarification questions more often. Fig \ref{fig:NetLogo Chat} and Fig \ref{fig:chatgpt} provide an exemplary comparison between NetLogo Chat and ChatGPT’s reaction to a simple modeling request. Here, ChatGPT immediately assumes details of the user’s needs and generates an entire model for the user to copy and paste. Whereas, NetLogo Chat attempts to first clarify the user’s needs by asking follow-up questions and suggesting exemplar answers. The suggestions in Fig \ref{fig:NetLogo Chat} serve as both an inspiration, in case learners get confused about what to write; and a shortcut, in case learners find any suggestions immediately usable.

For this feature to work effectively, it is essential to ask questions with quality. To achieve this, we used a few-shot approach and crafted templates for LLMs to follow. We conducted an informal evaluation of LLM's generated questions during our development process and empirical study. Across the board, the LLM we used was able to generate questions with acceptable quality, similar to the one demonstrated in Fig \ref{fig:NetLogo Chat}. A future design could embed a larger set of templates and retrieve a few relevant templates when needed.

\subsubsection{Invoke Authoritative Sources Whenever Possible}
\label{Authoritative Source}
Hallucination is another major concern for LLMs, particularly in an LRPL like NetLogo. For example, the code generated by ChatGPT in Fig \ref{fig:chatgpt} contains multiple syntax issues and requires human experts to address them. More powerful LLMs suffer from the same symptoms. We submitted similar sample requests to GPT-4, PaLM2, Anthropic Claude 2, and Falcon-180B: none was able to produce syntactically correct code for a classical NetLogo model. 

Following previous examples in related tasks\cite{joshi_repair_2023}, we integrated NetLogo's official documentation and model examples to help improve LLMs' and human performance. Different from previous studies, we not only provided related examples to LLMs, but also revealed them to users. By doing so, we seek to improve the transparency of LLM’s mechanism, foster trust in the LLM-driven system, and provide authoritative guides and examples for users even when LLMs might fail to provide precise support. 

\begin{figure}[h]
    \centering
    \includegraphics[width=240pt]{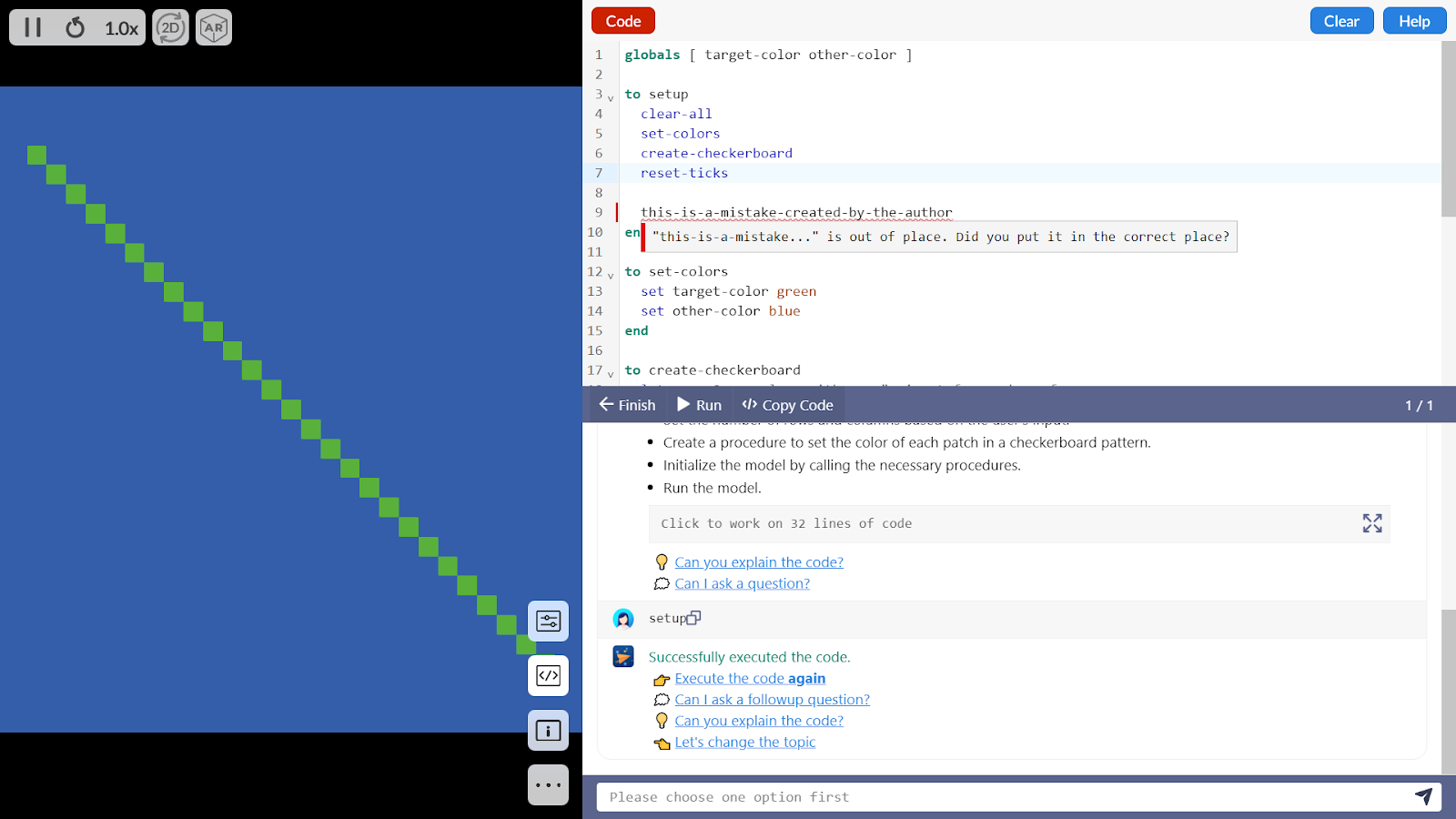}
    \caption{NetLogo Chat's embedded editor for generated code.}
    \label{fig:editor}
    \Description{This figure demonstrates the interface of Turtle Universe (a version of NetLogo). On the left, there is a visualization of a simple model that draws a diagonal line. On the top-right, there is a code editor that has the code and a mistake introduced by the researcher. The editor shows a linting message and an "explain" button. On the bottom-right is the conversation and interaction history between the user and NetLogo Chat.}
\end{figure}

\subsubsection{Integrate with the IDE and Enhance Troubleshooting}
\label{Troubleshooting}
We seek to integrate NetLogo Chat into NetLogo’s IDE beyond integrating a conversational assistant parallel to the code editor. To facilitate a constructionist learning experience, the code editor needs to be integrated into the conversational interface, where learners can work with smaller snippets of code with more ease. Thus, the design might lower the threshold for learners to tinker with the code, a key learning process advocated by the constructionist literature \cite{papert_situating_1991, turkle_epistemological_1990}. 

Fig \ref{fig:editor} provides a concrete example, where the embedded editor displays a piece of generated code. Instead of having to copy and paste the piece back into the main editor, the user could first see if any syntax issues exist in the code; run the code within a conversation; and ask follow-up questions or raise additional requests, before putting back a working code snippet into their projects.

To further support the user’s troubleshooting, in addition to error messages, NetLogo Chat will display extra debugging options for users. Users could choose to look for an explanation, or ask the LLM to attempt fixing the issue on its own, or with the user's ideas. During the process, the system will attempt to find documentation and related code examples to reduce hallucinations. Building on the literature on error messages' impact on learning\cite{becker2019compiler}, we also clarified many messages to provide a better context for humans and both LLM-based systems used in the study.

\subsection{Technical Implementation}
\label{Technical Implementation}
\begin{figure}[h]
    \centering
    \includegraphics[width=240pt]{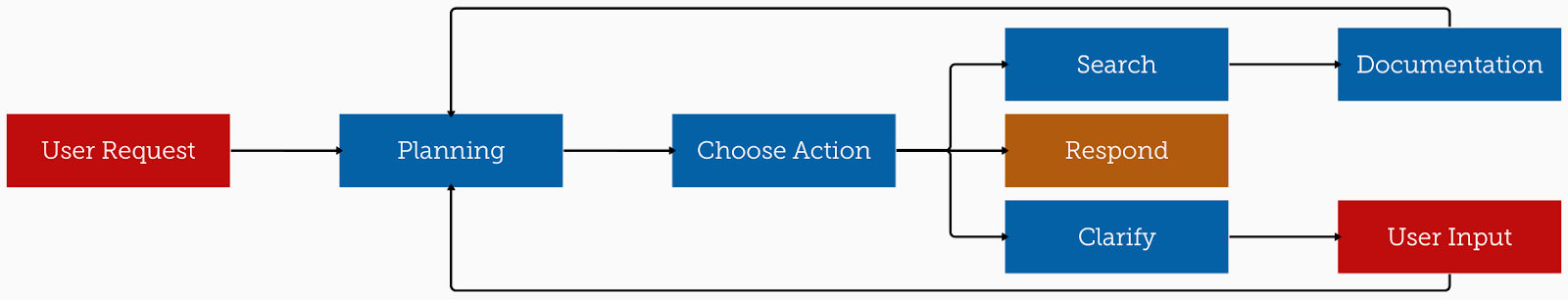}
    \caption{A brief outline for NetLogo Chat’s LLM workflow.}
    \label{fig:workflow}
    \Description{In this figure, from left to right: User Request => Planning => Choose Action => {Search => Documentation; Respond; Clarify => User Input} => Planning => ...}
\end{figure}

Since OpenAI started to provide fine-tuning on GPT-3.5-turbo (the version also used in ChatGPT Free) only after we concluded the main study in July, NetLogo Chat was implemented with prompt engineering techniques. We built our project on ReAct\cite{yao_react_2022}, a prompt-based framework that could reduce hallucination, improve human interpretability, and increase the trustworthiness of LLMs. By requiring LLMs to generate an action plan and delegate the action to a third-party conventional agent (e.g. search for documentation, ask clarification questions, conduct a static syntax check, etc.) before composing the final response, the framework provides a promising pathway to integrate external inputs (e.g. human input, official documentation) into LLM workflows. Fig \ref{fig:workflow} depicts a rough outline of NetLogo Chat’s workflow. Imagine a user requests to "create a predation model": 

\begin{enumerate}
  \item{The LLM is instructed, in the prompt, to first elaborate on the request (planning): "The user intends to create an agent-based biology model related to predation. However, it is unclear what exactly the user wants. We need to ask follow-up questions."}
  \item{Next, the LLM is instructed to choose an action from the list: Ask clarification question(s); Search for documentation; Write a response; Say sorry. Here, imagine the LLM chooses "Ask clarification question(s)" based on the planning.}
  \item{Then, the LLM needs to generate some questions based on the request. Because LLMs are trained on real-world data, it is not difficult for them to come up with some ideas. For example, "What species do you want to put in the model?" The LLM is also instructed to provide some examples, e.g. "Wolf", "Sheep".}
  \item{When the user replies to the questions, the loop restarts from step (1). Since there is sufficient information about the request, the LLM decides to search for information, and also generates keywords for the search, e.g. "Wolf-sheep predation model in NetLogo". }
  \item{The system conducts a semantic search on a pre-assembled database of NetLogo's official documentation and code examples. The system returns the search result, use it as a new round of input, and restarts from step (1).}
  \item{With inputs from both the user, who clarified the request; and the database, which supplies the example; the LLM plans again, chooses to write a response, and generates its final response.}
\end{enumerate}

In the example, we initiated three requests with the LLM, each with a prompt template that results in a structured response\cite{yao_react_2022} (e.g. any response needs to have a Plan, an Action, and a Parameter). Each request could use a different LLM that works best for the specific request. Using this approach, the system has the potential to balance cost, performance, speed, and privacy. For example, a future iteration of NetLogo Chat could leverage a fine-tuned local LLM to probe the user's intentions and search for documentation. Then, with any personal or sensitive information stripped away, the system could forward the compiled request to a powerful online LLM (e.g. GPT-4). 

For the empirical study, we chose GPT-3.5-turbo-0613 as NetLogo Chat’s LLM backend. First, we expect most participants to be using the free version of ChatGPT, driven by the same LLM. In this way, we would have a fair playing field for the empirical study, where both systems will be used. Second, at the time of our study, the response time for GPT-4 was too long to sustain a real-time experience, while we had no access to other NetLogo-capable LLMs’ APIs. Although we did observe some remarkable improvement when internally evaluating the system (e.g. ChatGPT has trouble answering questions for lesser-known NetLogo keywords, while NetLogo Chat does not), a more systematic evaluation rubric is needed for future research.

\section{Empirical Study}
\subsection{Participants}
For the empirical study, we recruited 30 adult participants through NetLogo’s official Twitter and mailing lists; and through the Complexity Explorer, a website run by Santa Fe Institute (SFI) to distribute learning resources of agent-based modeling (ABM). The exact breakdown of participants’ demographic data can be seen in Table \ref{tab:demographics}. The participant pool largely represented the scientific modeling community in NetLogo’s main audience, with a majority of participants coming from STEM disciplines. Many participants were also related to the educator sector. 6 participants (20\%) were instructors who teach or are interested in teaching NetLogo in classrooms; 4 (13\%) were graduate-level students interested in learning NetLogo, making up a third of the population. Participation in the study was voluntary. All participants signed an online consent form on Qualtrics. 

Building on the tradition of understanding the difference between experts and novices\cite{chi1981categorization}, we separated the participants into experts and novices using self-reported survey data. To mitigate the effect of inaccurate responses, NetLogo experts in the team, who have been core developers and instructors of NetLogo, watched every video and decided if a participant greatly overestimated or underestimated their capabilities. We considered the participant's discussions with the interviewer, the think-aloud process, and the coding behaviors. A vast majority of users' reports correspond with the experts' judgment. Then, to simplify the analysis, we separated participants (Table \ref{tab:participants}) by their levels into two main categories: experts, who are either experts in NetLogo or programming in general; and novices. In the study, we denote experts by the prefix E (E01-E17) and novices by N (N01-N13). 13 experts had previous experience with ChatGPT (76\%), including programming (65\%, n=11). 11 novices (85\%) also used ChatGPT before, but much less for programming (38\%, n=5).

\begin{table}[]
  \caption{Overview of Participant Demographics (n=30)}
  \label{tab:demographics}
\begin{tabular}{| p{40pt} | p{170pt} |}
\hline
Gender     & Females: 10 (33\%); Male: 19 (63\%); Non-binary: 1 (3\%)                                                          \\ \hline
Geography  & Africa: 1 (3\%); Asia and Oceania: 5 (17\%); Europe: 8 (27\%); Latin America: 2 (7\%); North America: 14 (47\%). \\ \hline
Occupation & Academics: 14 (47\%); Professionals: 12 (40\%); Students: 4 (13\%)                                               \\ \hline
\end{tabular}
\end{table}

\begin{table*}[]
  \caption{Participant Information}
  \label{tab:participants}
\begin{tabular}{|l|l|l|l|l|}
\hline
\textbf{ID} & \textbf{Region}  & \textbf{Level (NetLogo)} & \textbf{Level (Programming)} & \textbf{Occupation} \\ \hline
\rowcolor{Expert}
E01         & North America    & Expert                   & Expert                   & Professional        \\ \hline
\rowcolor{Expert}
E02         & Asia and Oceania & Expert                   & Intermediate             & Academic            \\ \hline
\rowcolor{Expert}
E03         & Latin America    & Intermediate             & Expert                   & Academic            \\ \hline
\rowcolor{Expert}
E04         & North America    & Expert                   & Expert                   & Academic            \\ \hline
\rowcolor{Expert}
E05         & Europe           & Intermediate             & Expert                   & Academic            \\ \hline
\rowcolor{Expert}
E06         & North America    & Intermediate             & Intermediate             & Academic            \\ \hline
\rowcolor{Expert}
E07         & Latin America    & Intermediate             & Intermediate             & Professional        \\ \hline
\rowcolor{Expert}
E08         & Asia and Oceania & Intermediate             & Intermediate             & Professional        \\ \hline
\rowcolor{Expert}
E09         & Asia and Oceania & Intermediate             & Expert                   & Professional        \\ \hline
\rowcolor{Expert}
E10         & North America    & Intermediate             & Intermediate             & Academic            \\ \hline
\rowcolor{Expert}
E11         & Africa           & Intermediate             & Expert                   & Academic            \\ \hline
\rowcolor{Expert}
E12         & North America    & Intermediate             & Intermediate             & Academic            \\ \hline
\rowcolor{Expert}
E13         & Europe           & Expert                   & Novice                   & Academic            \\ \hline
\rowcolor{Expert}
E14         & Europe           & Intermediate             & Intermediate             & Academic            \\ \hline
\rowcolor{Expert}
E15         & Asia and Oceania & Expert                   & Expert                   & Student             \\ \hline
\rowcolor{Expert}
E16         & Asia and Oceania & Novice                   & Expert                   & Professional        \\ \hline
\rowcolor{Expert}
E17         & Europe           & Intermediate             & Expert                   & Academic            \\ \hline
\rowcolor{Novice}
N01         & North America    & Novice                   & Novice                   & Professional        \\ \hline
\rowcolor{Novice}
N02         & North America    & Novice                   & Novice                   & Academic            \\ \hline
\rowcolor{Novice}
N03         & North America    & Novice                   & Novice                   & Professional        \\ \hline
\rowcolor{Novice}
N04         & North America    & Novice                   & Intermediate             & Student             \\ \hline
\rowcolor{Novice}
N05         & Europe           & Novice                   & Intermediate             & Student             \\ \hline
\rowcolor{Novice}
N06         & Europe           & Intermediate             & Novice                   & Student             \\ \hline
\rowcolor{Novice}
N07         & North America    & Novice                   & Intermediate             & Professional        \\ \hline
\rowcolor{Novice}
N08         & North America    & Novice                   & Intermediate             & Professional        \\ \hline
\rowcolor{Novice}
N09         & North America    & Novice                   & Novice                   & Professional        \\ \hline
\rowcolor{Novice}
N10         & North America    & Novice                   & Intermediate             & Professional        \\ \hline
\rowcolor{Novice}
N11         & Europe           & Novice                   & Intermediate             & Academic            \\ \hline
\rowcolor{Novice}
N12         & Europe           & Novice                   & Novice                   & Academic            \\ \hline
\rowcolor{Novice}
N13         & North America    & Intermediate             & Novice                   & Professional        \\ \hline
\end{tabular}
\end{table*}

\subsection{Interviews}
\label{Interviews}
Our study was conducted in 3 phases: 

\begin{enumerate}
  \item We pilot interviewed 3 experts invited from NetLogo’s online community. Each was asked to comment on LLMs for NetLogo learning, as well as on ChatGPT and an early prototype of NetLogo Chat. 
  \item We improved the design of NetLogo Chat based on what we learned from the pilot interviews and revised the interview protocol accordingly.
  \item We conducted formal interviews with 27 online participants (30 in total). 
\end{enumerate}

Each semi-structured interview lasted between 60-90 minutes and was video recorded. Prior to each formal interview, participants were asked to come up with a short NetLogo task that they were interested in working on. Almost every participant brought forward a modeling task from their career domain or personal interest, e.g. to model "how honeybees decide to regulate the temperature of the hive", or "the spread of conflicting ideas". Only once, when the task scope was too complicated for the session, did we ask the participant to bring another. During any part of the interview process, interviewers generally followed the protocol, asking follow-up questions when needed. Specifically:

\begin{enumerate}
  \item We asked baseline questions, e.g., “What do you think are the potential advantages / disadvantages of using LLMs in supporting your learning and programming of NetLogo?” (in 2 separate questions)
  \item We asked the participant to work on their task with the help of ChatGPT. Then, we asked the same baseline questions again, then asked “What do you like or dislike about the interface”. Repeat the procedure with NetLogo Chat;
  \item If time permitted, we further asked about their preferences for learning and/or programming with NetLogo and asked which feature they wanted to add/remove from either system. Here, the objective was not to strictly compare between the two systems, but to elicit more in-depth discussions over LLM-based interfaces.
\end{enumerate}

Since almost all users have already engaged with ChatGPT, we did not randomize the order of ChatGPT/NetLogo Chat. Also, 3 participants used the paid version (GPT-4) during the task with ChatGPT. While much of the generated data comes from the inevitable comparison between the two systems, we chose not to interpret them as objective comparisons. Instead, the different design principles underpinning the systems presented two objects to think with\cite{papert_mindstorms_1980}, that our participants drew on during their reflections and discussions of LLM-based programming interfaces.

\subsection{Data Analysis}
Our interviews resulted in around 40 hours of video data. Around half of our data is behavioral in nature, where participants worked on their tasks and were encouraged to think aloud; the other half is more verbal, where participants answered questions. As such, each interview was not only transcribed verbatim, but also watched by a researcher to create observational notes. The two streams were then combined into a single archive for analysis.

Based on our research questions, we iteratively applied the grounded theory approach\cite{corbin_grounded_1990} to analyze our data. During each step, the research team fully discussed the discrepancies between each researcher and iteratively refined the codebook to improve consistency. The analysis reached theoretical saturation at around 50\% of interviews, when additional interviews no longer revealed unexpected major insights for our research questions. Then, we finished the rest of qualitative coding with the finalized codebook (Table \ref{tab:codebook}).
\begin{enumerate}
  \item Four researchers open-coded 2 interviews, one from a novice and one from an expert, to summarize the topics mentioned by participants. During this process, researchers coded in different tabs to avoid interference. Three broad themes emerged from this phase: participants' approaches to programming; participants' interactions with AI systems; and their comments on AI systems. 
  \item Taking notes of the emerging themes, the first author created a preliminary codebook that categorizes dozens of codes into themes. Each researcher coded another 2 interviews in different tabs. In this phase, we refined the themes into approaches to programming (which also helps to separate experts and novices); perceptions and observed behaviors related to AI systems; and comments on AI systems' abilities. 
  \item Based on the coding results, the first author created a formal codebook, with definitions clarified based on the discrepancies between researchers (Table \ref{tab:codebook}). To reduce the unbalanced influence of subjective interpretation, researchers only coded explicit behaviors; or direct comments. To avoid missing insights, researchers were instructed to highlight places where existing codes are insufficient to cover the topics. During the first two weeks, a few codes were created or merged as a result of discussions. We retrospectively revised our coding.
\end{enumerate}

\begin{table*}[]
  \caption{An Overview of the Codebook}
  \label{tab:codebook}
\begin{tabular}{|l|p{350pt}|}
\hline
\textbf{Code}       & \textbf{Definition}                                                                                                                    \\ \hline
\textbf{Approaches} & User's perceptions about their approach to programming tasks, e.g. planning, separating into smaller pieces, or working on it as a whole. \\ \hline
Learning            & How users learn NetLogo or programming in general, or think that people should learn.                                                  \\ \hline
Coding              & How users organize or write their code, or think that people should organize or write.                                                 \\ \hline
Help-seeking        & How users seek help in general, or think that people should seek help.                                                                 \\ \hline
\textbf{Human-AI}   & User's perception and behaviors related to Human-AI relationship.                                                                      \\ \hline
Prior               & Users' prior experiences with ChatGPT or other AI-based interfaces.                                                                    \\ \hline
Attitude            & Users' attitudes toward AI in general, or specific AI-based systems.                                                                   \\ \hline
Effort              & AI's influence on how much, and what kind of, efforts that humans made or need to make.                                                 \\ \hline
\textbf{Abilities}  & User's perception related to AI's abilities.                                                                                           \\ \hline
Response            & AI's ability to provide desirable responses for humans.                                                                                \\ \hline
Support             & AI's ability to support learning/coding of NetLogo.                                                                                    \\ \hline
Interactivity       & AI's ability to facilitate helpful interactions with humans.                                                                            \\ \hline
\end{tabular}
\end{table*}

Based on the codebook, the first author iteratively incorporates themes into an outline.  To further mitigate individual differences, researchers were asked to include as many codes as possible for each quote or observation. 

\section{Findings}
\subsection{Perception: Before and After Interaction}
\begin{table*}[]
  \caption{Novices and Experts' Perceptions on LLM-based Interfaces for NetLogo}
  \label{tab:perception}
\begin{tabular}{| p{90pt} | >{\cellcolor{Expert}}p{150pt} | >{\cellcolor{Novice}}p{150pt} |}
\hline
                                                & \textbf{Experts}                                              & \textbf{Novices}                                                                              \\ \hline
                                                & LLMs could save human time and effort, especially in syntax.  & LLMs could save human time and effort, especially for syntax, and provide emotional benefits. \\ \cline{2-3} 
\multirow{-2}{*}{\textbf{Before, Positive}} &                                                               & LLMs could help troubleshooting.                                                              \\ \hline
                                                & LLMs could mislead humans to suboptimal directions.           & While LLMs may make mistakes, it is no worse than humans.                                     \\ \cline{2-3} 
                                                & LLMs could hinder learning processes.                         & LLMs may not understand human intentions.                                                     \\ \cline{2-3} 
\multirow{-3}{*}{\textbf{Before, Negative}} & LLMs could only work on smaller tasks.                        & LLMs' responses are difficult to understand.                                                  \\ \hline
                                                & LLMs supported learning or practicing by saving time.    & LLMs supported learning or practicing by saving time.                                    \\ \cline{2-3} 
\multirow{-2}{*}{\textbf{After Interaction}}           & Will continue to use LLMs for learning or practicing NetLogo. & Will seek alternative learning resources before continuing to use LLMs.                       \\ \hline
\end{tabular}
\end{table*}

\subsubsection{Before Interaction: Positive Expectations}
Prior to the tasks, both novices and experts had positive expectations of LLM-based interfaces for NetLogo, with novices holding higher expectations than experts. 

Both novices and experts expected LLM-based interfaces to save human time and support human effort, especially compared to other help-seeking activities. With LLMs, human time and energy could be liberated for more high-level tasks (\expert{12}, \novice{03}). Educators felt that LLMs could facilitate more efficient teaching, allowing students to \say{more complicated things with relative ease}, spiking \say{their imagination.} (\expert{02}) LLMs can also bring emotional benefits by reducing the fear of \say{bothering the teachers or the experts} (\expert{14}) or asking \say{stupid questions} (\novice{06}).

Most participants highlighted AI's potential to help them with NetLogo’s syntax. For most participants, NetLogo is not the main programming language they used. Before the advent of ChatGPT, \novice{06} felt that she needed to \say{recite the words (syntax of NetLogo)}. Yet, the need was eliminated when \say{AI can teach you very quickly}. Many experts also needed support, as NetLogo \say{has very strict syntax rules} (\expert{07}) which makes writing more difficult.

Novices, in particular, expected that AI could be helpful for troubleshooting. \novice{08}, for instance, felt that LLMs could help him through the troubleshooting process by describing \say{what I'm trying to do and get a snippet of code that helps get me past that block}. For novices without a background in programming, this future looks promising. \novice{12} is interested in the potential to \say{make programming more approachable to students}.

\subsubsection{Before Interaction: Negative Expectations}
Almost every participant expressed concerns or reservations about LLM-based interfaces. Yet, the concerns of novices and experts were conspicuously different. 

Experts focused on preserving human judgment. \expert{01} believed that AI should not \say{replace human judgment and ability}. Similarly, \expert{06} insisted that \say{(human) has to do the main thinking and ideas and all of that.} \expert{17} felt that humans cannot let AI \say{take over the main reasoning and emotions, the emotions intervening in the decisions.} Many educators were also \say{concerned about learning} (\expert{13}), fearing the tendency to \say{default to the AI system to come up with the answers instead of working through it ourselves} (\expert{12}). Many experts explicitly explained their rationales. For example, \expert{08} was concerned that \say{if a model points me to a suboptimal direction, I will have no idea, because I haven't considered alternative structure}. \expert{15} feared that relying on AI responses might \say{make your horizon narrow} because she would miss learning opportunities when browsing through the models library. For computational modeling, AI also might lack \say{in-depth knowledge in a specific field} to create an entire model (\expert{05}). As such, \expert{05} would only trust AI to \say{finish a specific task}.

Novices were more optimistic and more concerned with their capabilities of understanding AI’s responses or making AI understand them. For example, while \novice{04} thought \say{one of the hypothetical drawbacks} to LLMs being \say{confidently incorrect}, they added that \say{people are like this too}. On the other hand, \novice{03} feared that she would waste more time with AI if \say{it didn't understand me, or if I had difficulty expressing}. \novice{02} acknowledged that \say{there is a limitation to not knowing how to code (on how much AI could help).} Without knowledge of NetLogo, \novice{11} felt difficult to spot LLM-generated mistakes.

\subsubsection{After Interactions: Different Impacts of Hallucination}
All participants encountered AI hallucinations throughout the sessions. While some participants rated NetLogo Chat higher than ChatGPT’s free version, most participants had similar changes in perceptions: experts, in general, reported more benefits from LLMs than novices.

Some participants reported more positively about NetLogo Chat's capabilities. Several experts questioned ChatGPT’s training in NetLogo, yet they trusted more in NetLogo Chat, for it incorporates authoritative sources (see \ref{Authoritative Source}). \expert{16} believed that NetLogo Chat \say{understands your NetLogo syntax} and \say{the basic aspects of NetLogo}. \novice{02} thought NetLogo Chat still had bugs but was \say{much more informative and precise than ChatGPT.} As NetLogo Chat is designed to support troubleshooting (see \ref{Troubleshooting}), \expert{04} thought NetLogo Chat \say{was able to kind of do some better troubleshooting to a certain extent, for it clarifies error codes}. 

In both cases, experts understood hallucinations as an inevitable part of human-AI collaboration and reacted with more leniency. When \expert{03} first encountered an incorrect response, he exclaimed: \say{Very interesting! You're mistaken.} \expert{05} felt that LLMs helped him \say{finish most of the code}, though he still needed to \say{debug and see if the code makes sense logically.} As experts did not rely on LLMs to resolve issues but mostly leveraged them as a shortcut, \expert{06} stated that hallucinations were instances \say{where the programmer needs to use own experience and discretion}, as risks would escalate if one extrapolates \say{what ChatGPT provides you in a wrong manner}. 

Novices, on the other hand, reported more obstacles and frustration, as they relied more on LLMs for their tasks. \novice{07} emotionally responded to a hallucination that ChatGPT \say{apparently made that shit up}. \novice{01} had difficulties to \say{fix the bugs that were in it (the generated code).} \novice{08}'s session ended up \say{hitting a dead end}, with the frustration leading him to \say{go consult other resources}.

Most novices and experts still thought that LLM-based interfaces supported their learning or practicing by saving time. Even though \novice{03} had \say{low trust} in ChatGPT, she still felt more confident after collaboration, for it \say{narrowed down the stuff I have to figure out myself and has made me much faster already.} As an educator, \novice{12} felt that LLMs facilitated a constructionist learning experience in which \say{you’re being thrown into the culture and have to learn it on the fly.} \expert{13} thought he learned a syntax from ChatGPT that would \say{save me time in the future} and the learning process was \say{a lot faster than if I were doing it by hand}. 

As experts reported more perceived benefits, they predominantly intended to continue using LLM-based interfaces for NetLogo. After the task, \expert{11} felt confident that \say{I can write anything I want to write}. Yet, many novices, driven by their frustration with LLMs, sought alternative learning resources before considering a return. \novice{04}, for instance, had a 180-degree turn: expressing great hope before the tasks, they now inclined to \say{build more by myself with my own code, without AI.} \novice{13} thought that she would prefer to work with \say{someone who is familiar with the programming language} together with LLMs.

\subsection{The Behavioral Gap Between Novices and Experts}
\begin{table*}[]
  \caption{Novices and Experts' Behaviors During Human-AI Collaboration}
  \label{tab:behavioral}
\begin{tabular}{| p{90pt} | >{\cellcolor{Expert}}p{140pt} | >{\cellcolor{Novice}}p{160pt} |}
\hline
                                                  & \textbf{Experts}                                                     & \textbf{Novices}                                                                                                          \\ \hline
                                                  & Many start by asking LLMs for a smaller aspect of the task.          & Most start by asking LLMs to work on the entire task.                                                                     \\ \cline{2-3} 
\multirow{-2}{*}{\textbf{Planning \& Prompting}} & \textit{"NetLogo, I would like to spawn 50 turtles"}                 & \textit{"I want to use netlogo to help me model how honeybees regulate the temperature in their hive. What should I do?"} \\ \hline
                                                  & Focus more on the generated code.                                    & Focus more on the generated instructions.                                                                                 \\ \cline{2-3} 
\multirow{-2}{*}{\textbf{Evaluating}}             & \textit{"Talks too much. I want the code, not the explanation yet."} & \textit{"I am reading the text a little bit and it spits out a bunch of code. So it did give me steps, which is nice."}   \\ \hline
                                                  & Most selectively copy and paste code, or write code on their own.    & Most start by copying and pasting LLM-generated code.                                                                     \\ \cline{2-3} 
\multirow{-2}{*}{\textbf{Coding}}                 & \textit{"It'd be that I just take this and see what this does. "}    & \textit{“This time it gives me.. two boxes to copy.”}                                                                     \\ \hline
                                                  & Debug themselves, or with help from AI.                              & Debug with (more) help from AI.                                                                                           \\ \cline{2-3} 
\multirow{-2}{*}{\textbf{Debugging}}              & \textit{"Oh, I didn't ask him to move. That is my problem."}         & \textit{"I'm going to ask it the same question, but I'm confused why it said something about patches."}                   \\ \hline
\end{tabular}
\end{table*}

\subsubsection{Behavioral Gap in Planning and Prompting}
While experts’ and novices’ tasks were similar in terms of complexity, we observed differences between how novices and experts plan out their tasks. Since most participants gradually adapted their prompting styles, we focused on participants’ first-round prompts. 

Two initial prompting patterns, one emphasizing modeling the entire system and another focusing on smaller, initial aspects of the task, emerged from our interviews. Most novices adopted the first pattern (11/13, 85\%), while many experts adopted the second pattern (9/17, 53\%). Below, we introduce one vignette for each pattern:

\begin{enumerate}
  \item \novice{05} started by asking: \say{I need to make a model of the bunch of agents who are trying to promote political views to other people (...)}. Although he used GPT-4, the returned code still came with several syntax errors. \novice{05} then spent the next 20 minutes trying to ask GPT-4 to fix issues without success. He expected to \say{put the idea into it and we’ll run the code}, but in the end \say{it didn’t happen.}
  \item \expert{07} started by asking ChatGPT to \say{write code for drawing a rectangle}. When GPT-3.5 failed to further divide the rectangle, \expert{07} instantly pivoted to another strategy: \say{I have the following code that draws a rectangle. I want you to modify it so the rectangle is divided by two}. GPT-3.5 still failed, yet it produced working code and did \say{something close to it}.
\end{enumerate}

The second prompting pattern involved remarkable mental efforts to decompose and plan out the task. For example, \expert{07} described his approach as \say{separate into small, general tasks you want to do.} \expert{04} explained that he \say{just likes to iteratively build (the code)}. On the other hand, in the first pattern, many participants attempted to shortcut the efforts by delegating the tasks to AI, as \novice{05} said: \say{I just want to ask it (ChatGPT) to just directly make a code for this task and that's it.} 

By the end of the task, most participants had realized the importance of breaking tasks into smaller pieces for coding with AI. Naturally, when an LLM-based interface generated code with mistakes, a participant would be (implicitly) guided to ask smaller follow-up questions. Soon, many of them realized the benefits. \novice{01} thought it would be better if one \say{works through real small problems first, before getting to more complicated problems.} \novice{10} would \say{start with something really basic.} Experts using the first pattern had similar ideas. For example, \expert{12} decided to restart \say{with something simple and just work with it.}

\subsubsection{Behavioral Gap in Coding and Debugging}
As most participants engaged with an agent-based modeling task that they never worked on, both experts and novices learned some aspects of NetLogo with the help of AI - although, in different ways. Experts usually took a much more measured, prudent, and critical approach during coding and debugging, while novices mostly followed AI’s instructions.

Most novices focused on reading AI’s explanations and followed AI’s instructions during their coding processes. ChatGPT often gives instructions like \say{You can copy and paste this code into NetLogo and run it}. Even without this hint, almost all novices would copy and paste the generated code without much reading. The tendency worried some novices, but they had no choice: \say{I feel like I'm waiting for someone to tell me the answer, rather than learning how to solve it.} (\novice{11})

Experts put more emphasis on the code, often ignoring the explanations provided by AI. During their reading, experts evaluated and often criticized the responses, planning their next steps along the way. Only a few experts tried copying and pasting the code to see if they worked out of the box. Other experts selectively copied and pasted parts of the code into their programs, or wrote their programs with generated code on the side. Even when they copied and pasted the code, experts were more cautious. For example, while \expert{04} decided to \say{just take this and see what this does}, he also realized that AI-generated code would override his ideas and manually edited the code.

All participants inevitably had to debug parts of the generated code. Yet, novices sought support from AI more frequently and often struggled with AI responses. For example, \novice{12} would regularly \say{copy the code that doesn’t make sense and go back to AI to see if it can help me.} \novice{09} complained that while ChatGPT gave suggestions, \say{it obviously requires fiddling around with it.} As she had little idea about NetLogo, it became a purely trial-and-error experience. Even when AI did solve some errors, it was challenging for novices to learn from the process. For example, \novice{04} commented that while NetLogo Chat provided an automated process, it was still difficult for him to get the lesson, \say{since I didn’t write it myself.}

\subsubsection{Behind the Behavioral Gap: The Knowledge Gap}
We identified a knowledge gap that may lead to the behavioral gap. When novices realized that they needed to spend more effort decomposing the task or vetting AI responses, they found themselves lacking the necessary knowledge. We summarized, in participants’ own words, the four components of a knowledge gap that novices need to overcome when working with AI.

Novices reported the need for conceptual knowledge of modeling. For example, \novice{07} described his experience as \say{like being adrift on an ocean. Without a compass, and without a map.} With only a basic understanding of agent-based modeling, \novice{11} felt compelled to accept ChatGPT’s response as \say{I don't really know how to interpret some of the output from it.} Such feelings correspond with novices’ tendency to skim through AI responses. Whereas, some novices asked for help from LLMs with different degrees of success. \novice{04} first asked: \say{(...) Can you tell me what I will need to do before we begin?} With AI’s suggestions, \novice{04} had some more success asking follow-up questions. 

The unfamiliarity with the basic concepts of NetLogo and/or coding in general further adds to the difficulty in prompting and understanding. After reading a guide suggested by NetLogo Chat, \novice{07} realized that he \say{probably wouldn’t have chosen NetLogo to ever begin with} for his database-related task. Other novices were often confused by NetLogo’s terms, even when they were mostly in plain English. \novice{03} was confused about \say{why (ChatGPT) said something about patches} (note: patches are static agents that form NetLogo’s modeling world), and that deepened her reliance on ChatGPT. \novice{10} realized that she \say{only understand 20\% of what I am reading, so I can't vet it myself.} When the interviewer asked about adding comments into code, \novice{03} replied that while it might be helpful, she was still missing \say{the high-level understanding of how it comes together.}

Many novices also lack the experience for debugging, leading to more unsuccessful attempts and more frustrations. Participants, in particular novices, were often confused by error messages from NetLogo. \novice{01} acknowledged that \say{without background knowledge, it is hard to figure out what the bugs are, if (LLM) gives you information that is inaccurate.} Without experience in debugging, many novices felt frustrated and helpless as previously reported. On the other hand, \expert{12} noted that his students \say{might not be comfortable with the idea that debugging is a normal part of the process.} \expert{01} believed that \say{the user needs a little practice in debugging their own code} before working with LLM-based interfaces. 

Most novices felt a need to learn to interact with LLMs. After repeated failures, \novice{01} felt that he did not \say{even know what questions to ask to get it to, because it is not doing the right thing.} \novice{06} thought AI would help a lot if she could \say{learn more about how to use AI.} \novice{05} realized that he needed to use the correct keywords, for otherwise it \say{will never generate a good model.} This knowledge is relatively easier to acquire though: while \novice{09} felt that \say{how to ask questions is very important}, she believed that \say{you learn by actually doing it.}

\subsection{Needs for Guidance, Personalization, and Integration}
\begin{table}[]
  \caption{Users' Needs for LLMs: Guidance, Personalization, and Integration}
  \label{tab:needs}
\begin{tabular}{|
p{70pt} |
p{70pt} |
p{70pt} |}
\hline
\textbf{Guidance}                                                                                                & \textbf{Personalization}                                                       & \textbf{Integration}                                                    \\ \hline
Should provide clear, less technical responses, stay on topic, and give smaller pieces of information at a time. & Should provide responses based on users' preferred styles.                     & Should provide better support for coding chunks and iterative modeling. \\ \hline
Should provide responses based on authoritative sources and in NetLogo's language.                               & Should provide responses based on the knowledge levels and interests of users. & Should support working on existing modeling code.                       \\ \hline
Should assume less, clarify more, and stick to user intentions for modeling.                                     & Should support human help-seeking preferences in different ways.               & Should support input and output of computational modeling.              \\ \hline
\end{tabular}
\end{table}

\subsubsection{"Good" Responses, "Bad" Responses}
Participants generally appreciate and expect less technical, clear instructions. Many of them appreciate NetLogo Chat's design decisions that include authoritative sources in responses (see \ref{Authoritative Source}) and ask back clarification questions (see \ref{Program}). However, participants' preferences are also highly personal and situational. 

For both designs, some participants explicitly went against excessive or unnecessary explanations, particularly when the goal is primarily to accomplish a task at hand. For instance, \expert{09} complained that GPT-4 \say{talks too much. I want the code, not the explanation yet.} \expert{14} complained that while related code samples provided by NetLogo Chat could \say{contain a lot of good suggestions}, she wanted to move them to \say{another box or an expandable line}.

Some participants appreciated and hoped that LLMs could stay on topic and give smaller pieces of information at a time. \expert{01} thought NetLogo Chat would be more helpful if it only attempted to solve a bug \say{one at a time}, for users \say{always overfill their buffer}. Novices, in particular, prefer concrete, step-by-step responses, given the focus they put on AI-generated instructions. \novice{04} wanted to \say{test one by one if (LLM) gave me multiple suggestions.} Going beyond text responses, \novice{03} hoped that there could be \say{a visual to help me better understand, or internalize what different elements of the code are}, so her learning could move to a higher-level understanding of the code’s intention.

For NetLogo Chat, most participants reacted positively to the reference to authoritative sources (see \ref{Authoritative Source}), the usage of NetLogo's language, and the provision of links to sources. \expert{03} believed that \say{the possibility to go directly from this AI to the documentation} would be helpful for his students. \novice{10} \say{automatically like (NetLogo Chat's response) better} because it used \say{NetLogo's kind of turtle and patch language.} \expert{12} felt \say{a little bit more confident in the information I was getting because it seemed to be coming from inside of the application.} However, sticking too much to authoritative explanations might have a downside. \expert{10} complained that NetLogo Chat gave him \say{dictionary reference}, and \say{dictionary definitions are not especially helpful.}

Many participants, in particular experts, reacted positively when NetLogo Chat assumed less about and stuck more to their intentions (e.g. asking questions back, see \ref{Program}). For example, \expert{09} commented that ChatGPT (GPT-4) \say{assumed what I wanted it to do, whereas this one makes you specify your assumptions.} He prefers NetLogo Chat’s approach, because \say{it makes you think about the code more.} \expert{12} felt that NetLogo Chat’s clarification of intention was akin to \say{progressively guiding me towards a better prompt.} As transparency is a key factor in computational modeling, \novice{11} feared that if \say{anyone can produce an agent-based model, but without actually understanding all the parameters}, hidden assumptions introduced by ChatGPT could be detrimental.

\subsubsection{Need for Personalization}
In this section, we break down the strong needs of experts and novices for more personalization, besides response styles, into two themes: knowledge levels and help-seeking needs. 

Novices, in particular, felt a strong need for LLM-based interfaces to acknowledge their knowledge levels and produce responses accordingly. \novice{07} gave a stringent critique of both systems, feeling both systems were \say{not useful at all}, for both \say{presumes you know something about NetLogo}. \novice{08} felt that \say{ChatGPT has no idea of how much or how little I know about how to code in NetLogo, or how to code in general.} Solving this issue would require more personalized approaches. Coming from an educational background, both \novice{02} and \expert{03} suggested that LLMs should first probe the knowledge level of users before providing answers. 

Participants gave a variety of suggestions that were at times conflicting: 

\begin{enumerate}
  \item \textbf{Some participants prefer a guided walkthrough.} \novice{08} hoped that LLMs could walk him through the process and provide starting points. Both \expert{14} and \novice{03} hoped that LLMs could be used alongside video tutorials, where they could first see a successful example of human-AI collaboration and then ask follow-up questions. 
  \item \textbf{Some participants prefer contextual recommendations.} \novice{11} hoped that LLMs could show related code examples and provide \say{two or three other ways that you might look with}. \expert{10} suggested that LLMs provide in-context explanations if \say{you don’t remember the definition or explanation of a particular command}. 
  \item \textbf{Some participants hope that LLMs could support help-seeking from humans.} \expert{01} hoped that LLMs could help novices \say{explain my situation so that I can paste it to the user group}, so human experts could intervene more easily when AI fails to unstuck novices. Similarly, \expert{17} suggested that AI could be combined with \say{peer to peer answers and collaboration}.
  \item \textbf{Some participants believed that incorrect responses could become a learning opportunity.} \expert{02} was concerned that students might be \say{exposed to fewer options} with AI, compared with \say{coding from scratch}. \expert{03} feared that a system capable of directly producing solutions might deprive students of the debugging process, where they would have learned.” Novices also had similar feelings. After many hallucinated responses, \novice{08} thought that ChatGPT \say{forces me to learn as opposed to just getting code that’s ready to go.} To fully transform the moment of mistake into learning opportunities, educators suggest the design not to frame mistakes as failures, but rather \say{as a learning moment} (\expert{12}).
\end{enumerate}

\subsubsection{Need for Integration}
Compared with ChatGPT in a separate browser window, most participants appreciated the NetLogo Chat interface being an integrated part of the modeling environment. They are particularly in favor of the deep integration in NetLogo Chat's design that goes beyond placing a CA and an IDE side-by-side. We further identified many participants' need for a deeper integration.

Many participants appreciated the integration of a sandbox-like code editor in NetLogo Chat, where they can tinker with smaller, AI-generated code chunks and execute them on the fly (see \ref{Troubleshooting}). \novice{12} \say{definitely liked this feature of being able to go easily between the code and see what was changed and what was added.} \novice{04} appreciated that one can \say{see the code run} in the NetLogo IDE, which ChatGPT could not do. \novice{13} thought while some code generated by ChatGPT was \say{so comprehensive}, NetLogo Chat was able to break it down and make them \say{more conducive}. Participants also expressed further needs for iterative modeling. \expert{13} hoped that NetLogo Chat could help him \say{modularize all of my commands} by splitting the code into many smaller, more manageable chunks. \expert{12} asked for a comparison feature between versions of code chunks that could help him \say{iterative changes quickly}. 

In addition, participants also hoped that LLMs could help them reflect on longer pieces of (existing) code. \novice{02} and \expert{02} wanted AI to support the combination of multiple, smaller code chunks into a single, coherent code. As such, LLM-based interfaces should be able to work with longer pieces of code. Both \novice{06} and \expert{08} hoped that NetLogo Chat could \say{look at my code and make suggestions based on my code}. 

Many participants needed adaptive support for modeling more than just coding. Many requested AI support in building model interfaces that could be used to take in inputs or send out outputs. For example, \novice{06} needed NetLogo for her academic paper, hence plotting became \say{very important}. For educators like \expert{13}, while the canvas output was \say{good for the three-quarters of a project}, it hid \say{the real power of agent-based modeling - tracking the emergent properties of the model, rather than simply making bits run around the screen.} During the modeling processes, many interface parts could become necessary or unnecessary depending on situational needs. Integrated LLM-based interfaces need to go beyond a \say{side chat window} and support various spatial configurations for advanced users to decide on.

\section{Discussions}
Our study first reported, in detail, how novices and experts perceive and use LLM-based interfaces (ChatGPT \& NetLogo Chat) differently to support their learning and practice of computational modeling in an open-ended setting. Most participants appreciated the design direction NetLogo Chat is heading toward. However, they also expressed their needs for improved guidance, personalization, and integration which opens up huge design spaces for future improvement. 

\subsection{Guidance: Bridging the Novice-Expert Gap}
\begin{figure}[h]
    \centering
    \includegraphics[width=240pt]{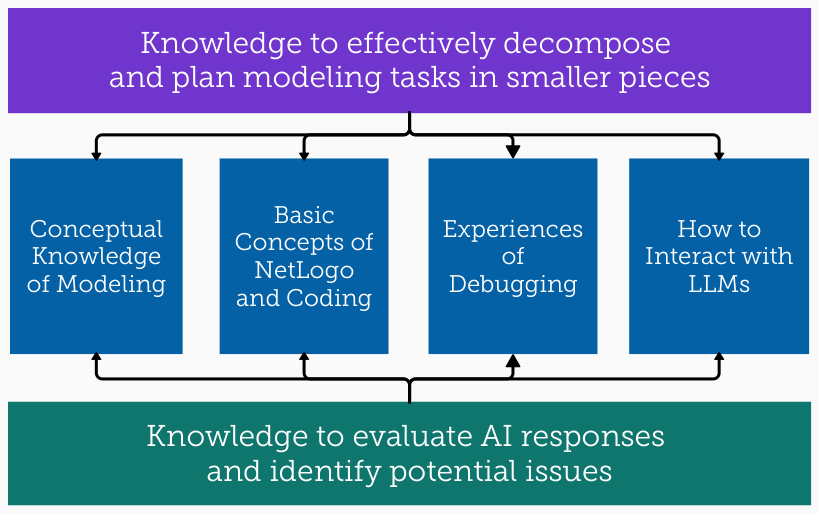}
    \caption{A preliminary theorization of the novice-expert knowledge gap.}
    \label{fig:knowledge}
    \Description{This figure summarizes our theorization of the gap. The gap has two parts: knowledge to effectively decompose and plan modeling tasks in smaller pieces; knowledge to evaluate AI responses and identify potential issues. Both comprise two parts: conceptual knowledge of modeling; basic concepts of NetLogo and coding; experiences of debugging; and how to interact with LLMs.}
\end{figure}

For most participants, guidance is what they need most from LLMs in programming. While \textbf{hallucinations} from LLMs constantly present a challenge to everyone, with a higher frequency to evaluate and debug AI responses, experts suffered less negative impact than novices. As a result, experts reported higher levels of perceived gains and more optimistic adoption plans than novices. While novices in our study also attempt to evaluate and debug AI responses, they are ill-equipped for these tasks. Without understanding the knowledge gap between experts and novices, it becomes impossible to design effective guidance.

Based on our empirical findings, we theorize the two types of knowledge novices might need when collaborating with AI in computational modeling (Fig \ref{fig:knowledge}). First, the knowledge to effectively decompose and plan modeling tasks. Second, the knowledge to evaluate AI responses and identify potential issues. We further identified four components of knowledge that both novices and experts reported to be essential: conceptual knowledge of modeling; basic concepts of NetLogo and coding; experiences of debugging; and how to interact with LLMs. To mitigate the impact of currently inevitable hallucinations of LLMs, it is essential to help novices get over the knowledge gap.

We propose three learning moments where design intervention might work best. \textbf{The first moment} is when users plan their next steps. While most novices started by delegating the planning process to AI, most of them eventually planned on their own. Here, we follow the constructionist learning theory for a broader understanding of planning that includes both rigid, formal plans and "softer", ad-hoc exploration of problem spaces\cite{turkle_epistemological_1990}. Both planning styles should be recognized as legitimate in learning and supported by the design \cite{turkle_epistemological_1990}. With our current design, most novices reported positive feelings when NetLogo Chat attempted to clarify their intentions and produce a plan for their task. Since this phase does not involve any generated code, more support could be provided, as novices may have fewer problems reading and evaluating natural language responses. They may also feel more comfortable asking questions about modeling or programming ideas, relating them to the generated code later, without fearing that they cannot (yet) read or write code. Moreover, LLMs could expand learners' visions by suggesting new ideas, proposing new plans, or taking notes of human ideas. When novices are confused about basic concepts, LLMs could suggest video or textual tutorials and provide Q\&A along the way. 

\textbf{The second moment} is when users read and evaluate LLM-generated code. Reading and understanding code is one of the most important aspects of computing education\cite{lopez2008relationships}. However, novices in our study were neither confident nor equipped for reading code. As a result, they intended to skip the code section. As predicted by the interest development framework\cite{michaelis_interest_2022}, the lack of skills (knowledge) and confidence (identity) may mutually enhance each other. Breaking the feedback loop requires designers to scaffold their reading experiences in both directions. By making explanations within code (as comments or tooltips) or visualizing the code structures (e.g. \cite{sorva2013review}), we might be able to help build novices’ connections between code syntax and real-world meanings. To build up learners' confidence, LLMs should deliver code pieces and explanations in adaptive sizes that work for learners. For learners who still could not succeed, the interface should further provide ad-hoc support that helps novices ask follow-up questions, or lead them to appropriate learning resources.

\textbf{The third moment} is when users need to debug their code. Debugging is considered a rich learning opportunity in constructionist learning\cite{kafai_turning_2020}. However, it is often associated with negative feelings that both manifested in prior literature\cite{whalley2021novice}, as well as our findings. Unfortunately, cognitive science has found that negative moods may further impede debugging performance\cite{khan2011moods}, enlarging the gap between novices and experts. Following the suggestions of educators in our study, we suggest that LLM-based interfaces could frame bugs in a more positive light, while providing a link to a successful human-AI collaborative debugging process for first-time learners. Both novices’ and LLMs’ debugging processes are often stuck in loops\cite{whalley_novice_2021, wu_rustgen_2023}. While such situations are inevitable, some expert participants suggest that LLM-based interfaces could encourage learners to seek help from another human. Help-seeking is recognized as an important part of programming education, yet novices often struggle with it\cite{marwan2020unproductive}. In such cases, LLM-based interfaces should further help them frame questions for human experts.

\subsection{Personalization: Beyond “Correctness” of LLMs}
Personalization has been identified as an essential factor for perceived autonomy when users interact with conversational agents\cite{yang_designing_2021}, for emotional and relational connections\cite{wellner_ihde_2023}, and for various educational benefits\cite{bernacki_systematic_2021}. Adding to previous literature, we found personalization to be a crucial factor for LLMs to facilitate effective guidance for learning, as participants expect LLMs to recognize their knowledge levels and react accordingly. 

While LLMs might have the potential to further the personalization of learning, recent research in LLMs focused on the “objective” capabilities, ignoring the personalized aspect of its evaluation. For example, technical reports of LLMs all reported benchmarks in whether they could produce functionally correct programs (HumanEval)\cite{chen_evaluating_2021}; if they could correctly answer multi-choice questions (MMLU)\cite{hendrycks_measuring_2020}; or if they could produce the correct answer of grade school mathematical problems (GSM-8K)\cite{cobbe_training_2021}. While working toward such “correctness” benchmarks is certainly crucial for LLMs to reduce hallucination and produce better responses, it becomes problematic when the definition of “helpfulness” or “harmfulness” is measured with a ubiquitous scale without individual differences \cite{bai_training_2022}, and such a definition has since been adopted by all major players in LLMs. 

At least in learning and practice programming, we argue that helpfulness cannot be a singular metric, but instead varies based on many factors. Corroborating with constructionist design principles\cite{resnick_reflections_2005}, we identified some potentially important factors such as knowledge levels and help-seeking preferences, while other factors such as culture, ethnicity, and gender could be as important. To support human learning, the full potential of LLMs could only be achieved through the recognition of epistemological pluralism\cite{turkle_epistemological_1990}: humans have different approaches toward learning, and technology needs to be tailored to human needs.

Most participants in our study expected or asked for personalization, in the sense that LLMs recognize their knowledge levels and help-seeking needs, yet today’s designs are still far from that. While it is virtually impossible to fine-tune thousands of LLM variants, LLMs’ role-play capabilities and novel prompt-based workflows (e.g. the one used by NetLogo Chat, or the concept of GPTs very recently released by OpenAI) have shown promising potential. As personalization requires the inevitable and sometimes controversial collection of user data, we suggest a more upfront approach: only collecting data that directly contributes to a more helpful AI (e.g. the knowledge level), only using data for this purpose, and explaining the benefits, risks, and privacy processes at the beginning. Alternatively, designers could also consider flowing the pathway of cognitive modeling, which deduces learners' knowledge levels from known interactions with the system\cite{sun2008introduction}. On the other hand, our understanding of users’ perceptions, behaviors, and needs for LLM-based programming interfaces has just begun, and we call on more studies to pursue this direction.

\subsection{Integration: LLMs for Computational Modeling}
For most participants, integration between LLM-based interfaces and modeling environments goes beyond stitching a chat window into the IDE. While most of them appreciated NetLogo Chat’s design directions, they put forward many needs that are worth considering in future design. Here, we briefly discuss the two major themes: support for troubleshooting; and support for modeling. For troubleshooting:

\begin{enumerate}
  \item \textbf{The capability to work on smaller snippets of code, with the capability to execute, explain, and debug code in context.} For both humans and LLMs\cite{hou_large_2023}, debugging complicated code is known to be difficult. NetLogo Chat has made the first step in reducing the scope to smaller code chunks. As such, it becomes easier for both humans to debug and LLMs to support their debugging processes. Whereas, more work is needed to bring together the code chunks into coherent full programs.
  \item \textbf{The capability to leverage authoritative NetLogo documentation in generated responses, as well as for the user’s own reference.} In debugging contexts, LLMs’ tendency to hallucinate becomes more frustrating. By providing users and LLMs with authoritative explanations within the debugging context, NetLogo Chat may reduce the effort for users to seek related information, which is also known to be difficult for novices\cite{dorn_lost_2013}. More work is needed to explain in a more personalized way: for example, pure novices may need explanations for every basic term.
  \item \textbf{The capability to automatically send in contextual information (i.e. code and error messages) for LLM to troubleshoot.} Users generally appreciated NetLogo Chat's design decision to support troubleshooting. However, the convenience came with a potential price: when using NetLogo Chat, users were more likely to ask LLMs for help, which might lead to fewer human attempts and learning opportunities. Further studies are needed to understand this design balance better.
\end{enumerate}

Many participants also asked for features that specifically support their computational modeling tasks, which are known to have different priorities from programming in general\cite{pylyshyn_computational_1978}. Here, two more capabilities are warranted:

\begin{enumerate}
  \item \textbf{The capability to assume less, actively probe, and stick to user intentions.} In addition to the potential learning opportunities (see Discussion 1), for participants, hidden assumptions in scientific modeling are particularly harmful. While users appreciate NetLogo Chat's direction in having LLMs ask questions back, future interfaces should be able to facilitate the conversational build-up of plans and steps, further supporting users to program computers piece-by-piece rather than falling to hidden assumptions made by LLMs. 
  \item \textbf{The capability to support modeling practices beyond coding.} Building the program is only one step; computational modeling also involves design, data visualization, and validation\cite{weintrop_defining_2016}. For LLM-based interfaces to support modeling practices, future interfaces should go beyond coding to support users’ efforts throughout the modeling process. 
\end{enumerate}

\section{Limitations and Future Work}
There are limitations to our study that warrant future work. As a widely used agent-based modeling language, a deeper understanding of user perceptions, behaviors, and needs for LLM-based interfaces around NetLogo may inform us of design choices for other modeling environments. Future work should consider computational modeling or programming environments that might have different priorities. Since the NetLogo language was designed for an audience without a computer science background\cite{tisue_netlogo_2004}, it becomes more important and meaningful to understand how to design for bridging the novice-expert gap in LLM-based interfaces. However, it is unclear whether our findings and suggestions would sufficiently support novices' and experts’ learning and practice of NetLogo. Using a more rigid rubric to distinguish between experts and novices might improve the rigor of our study. A quantitative, controlled study in the future might further (in)validate our findings and suggestions. As such, we plan to work on a new iteration of NetLogo Chat design and empirical study to fully understand the design implications.

Although we aimed to recruit participants representative of NetLogo’s global audience, our participant pool was not as representative as we hoped in two key dimensions. First, our participants were mostly professionals, academics, and graduate students. While K-12 teachers and learners are another major audience for NetLogo and agent-based modeling and may have different priorities and preferences\cite{sengupta_programming_2015}, only one K-12 teacher was present in the study. More studies are warranted to further the empirical understanding of LLM-based interfaces in education contexts. Second, the demographics of our participants skewed towards North American and European, highly educated, and male. Such a group of participants, recruited voluntarily, might manifest higher than average acceptability toward novel technology, e.g. most of our participants have already engaged with ChatGPT. For future work, researchers need to recruit a more balanced and diverse group of participants, if the goal is for LLM-based programming interfaces to equitably support novices and experts throughout the world. 

\section{Conclusion}
As Large language models (LLMs) have the potential to fundamentally change how people learn and practice computational modeling and programming in general, it is crucial that we gain a deeper understanding of users' perceptions, behaviors, and needs in a more naturalistic setting. For this purpose, we designed and developed NetLogo Chat, a novel LLM-based system that supports and integrates with a version of NetLogo IDE. We conducted an interview study with 30 adult participants to understand how they perceived, collaborated with, and asked for LLM-based interfaces for learning and practice of NetLogo. Consistent with previous studies, experts reported more perceived benefits than novices. We found remarkable differences between novices and experts in their perceptions, behaviors, and needs. We identified a knowledge gap that might have contributed to the differences. We proposed design recommendations around participants' main needs: guidance, personalization, and integration. Our findings inform future design of LLM-based programming interfaces, especially for computational modeling.

\begin{acks}
We would like to express our gratitude to the \href{https://community.netlogo.org/}{NetLogo Online community} and \href{https://www.complexityexplorer.org/}{Complexity Explorer} for their help and support. We are especially thankful to the hundreds of NetLogo users who volunteered for the study. We would also like to thank current and former members of our lab and anonymous youth users of Turtle Universe, who provided valuable feedback and ideas during our design process. Specifically, we want to acknowledge the intellectual contributions of Umit Aslan; Aaron Brandes; Jeremy Baker; Jason Bertsche; Matthew Berland; Sharona Levy; Jacob Kelter; Leif Rasmussen; David Weintrop; and Lexie Zhao. Finally, we appreciate the valuable and actionable feedback from our anonymous CHI reviewers, which significantly strengthened the paper. 
\end{acks}

\bibliographystyle{ACM-Reference-Format}
\bibliography{bibliography}

\end{document}